\documentclass[useAMS,usenatbib,usegraphicx]{mn2e}
\usepackage{epsf}
\usepackage{amssymb}

\title[The impact of lens galaxy environments on the image separation
distribution] 
{The impact of lens galaxy environments on the image separation distribution}
\author[M. Oguri, C. R. Keeton \& N. Dalal]
{Masamune Oguri,$^{1, 2}$\thanks{E-mail: oguri@astro.princeton.edu} 
Charles R. Keeton$^3$
and Neal Dalal$^4\thanks{Hubble Fellow}$\\
$^1$Princeton University Observatory, Peyton Hall,
Princeton, NJ 08544, USA.\\
$^2$Department of Physics, University of Tokyo, Hongo
7-3-1, Bunkyo-ku, Tokyo 113-0033, Japan.\\
$^3$Department of Physics and Astronomy, Rutgers University, 136
Frelinghuysen Road, Piscataway, NJ 08854 USA.\\
$^4$Institute for Advanced Study, Einstein Drive, Princeton NJ 08540,
USA.
} 
\begin{document}

\date{\today}

\voffset- .65in

\pagerange{\pageref{firstpage}--\pageref{lastpage}} \pubyear{2005}

\maketitle

\label{firstpage}

\begin{abstract}
We study the impact of lens galaxy environments on the image separation
distribution of lensed quasars. We account for both environmental
convergence and shear, using a joint distribution derived from galaxy
formation models calibrated by galaxy--galaxy lensing data and number
counts of massive elliptical galaxies.  We find
that the external field enhances lensing probabilities, particularly
at large image separations; the increase is $\sim$30\% at $\theta=3''$
and $\sim$200\% at $\theta=5''$, when we adopt a power-law source
luminosity function $\Phi(L)\propto L^{-2.1}$. The enhancement is
mainly driven by convergence, which boosts both the image separation
and magnification bias (for a fixed lens galaxy mass).  These effects
have been neglected in previous studies of lens statistics.  Turning
the problem around, we derive the posterior convergence and shear
distributions and point out that they are strong functions of image
separation; lens systems with larger image separations are more likely
to lie in dense environments.
\end{abstract}


\section{Introduction}

Strong gravitational lensing offers a unique probe of cosmology and
the physical properties of early-type galaxies. For example, the
ensemble properties of a well-defined sample of strong lens systems
can constrain the cosmological constant \citep{turner90,fukugita90,
kochanek96b,chae03a,mitchell05}, the density profile and evolution
of early-type galaxies \citep{kochanek00,rusin03a,rusin03b,rusin05}, and the
amount of substructure in lens galaxies \citep{metcalf01,dalal02,kochanek04}. 
In addition, the image separation distribution of lensed quasars
\citep{turner84} probes the efficiency of baryon cooling inside dark
matter halos of different masses, and thereby constrains models of
galaxy formation \citep{kochanek01,keeton01a,oguri02}.  So far, more
than 80 lens systems have been discovered, and the largest homogeneous
subsample suitable for statistical studies is the Cosmic Lens All-Sky
Survey (CLASS), containing 13 lensed radio sources at image separations
$0\farcs3<\theta<3''$ \citep{myers03,browne03}. A larger statistical
sample is being constructed from the Sloan Digital Sky Survey (SDSS),
which has already discovered more than 10 new lenses including the
largest-separation lensed quasar known to date \citep{inada03,oguri04}. 

Lensing probability distributions are often computed using a clean
spherical lens object without external fields. However, in reality, 
most observed lenses have non-spherical galaxies, and many require a
significant external tidal shear \citep{keeton97}. While a portion
of the shear may come from fluctuations of matter along the line of
sight \citep{seljak94,barkana96,momcheva05}, much of it is thought to
be associated with mass in the immediate environment of the lens
galaxy. Most lens galaxies are early-type, which lie preferentially
in dense environments. For instance, 
\citet[][see also \citealt{blandford01}]{keeton00} used galaxy
demographics (specifically, the galaxy luminosity function as a
function of type and environment) to predict that at least $\sim$25\%
of lens galaxies lie in groups or clusters of galaxies.  Indeed, more
than 20 lens galaxies are already known to or suspected to be surrounded
by groups or clusters \citep{young81,kundic97a,kundic97b,tonry98,
fischer98,tonry99,tonry00,fassnacht99,hagen00,fassnacht04,kneib00,soucail01,faure02,faure04,fassnacht02,johnston03,inada03,morgan05,oguri05,mckean05,momcheva05}.

Conventional wisdom holds that ellipticity and environment mainly
affect the relative numbers of double and quadruple lenses and do
not significantly modify the total lensing probability
\citep[][but see \citealt{keeton04}]{kochanek96b,keeton97,rusin01,chae03a}.
\citet{huterer05} showed explicitly that ellipticity and shear
change the lensing probability by only a few percent for most
source luminosity functions \citep[see also][]{premadi04}. They
also found that shear (and to a lesser extent ellipticity) shifts
and broadens the distribution of image separations for a given lens
galaxy.  However, they did not put the two pieces together and
discuss changes in the overall image separation distribution.  Also,
\citeauthor{huterer05} (and everyone else, for that matter) neglected
the effects of external convergence from the environment.  External
convergence is often omitted from models of individual lenses because
the mass sheet degeneracy renders it unmeasurable \citep{gorenstein88,saha00}.
Nevertheless, theoretical studies show that external convergence
affects lensing analyses in many important ways \citep{keeton04}.

\begin{figure}
\begin{center}
 \includegraphics[width=1.0\hsize]{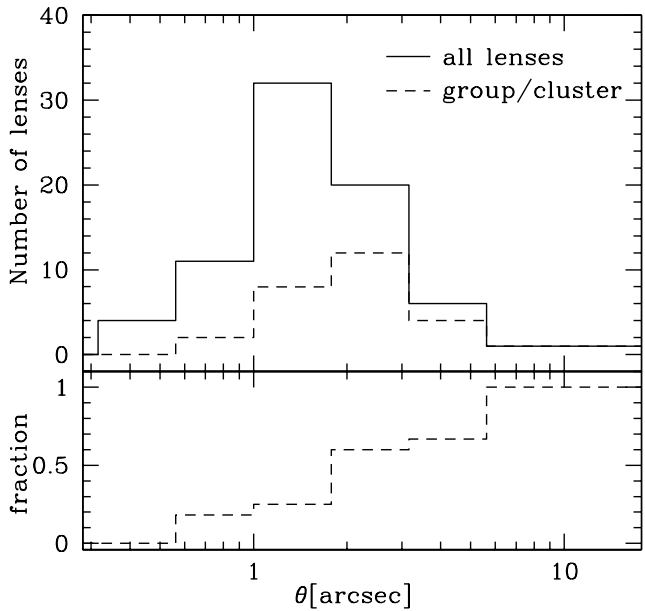}
\end{center}
\caption{Top: Image separation distributions of all lensed quasars
 ({\it solid}) and those whose lens galaxies are confirmed to or
 suspected to lie in a group or cluster ({\it dashed}). Note that not
 all lensed quasars have had their environments characterized
 observationally.  Bottom: The fraction of lensed quasars in groups and
 clusters as a function of image separation. 
\label{fig:sepdist_cas}}
\end{figure}

There is already some observational evidence that lens galaxy
environments affect the image separation distribution.
\citet{oguri05} found an overabundance of lensed quasars with image
separations $\theta\sim 3''$. They argued that the excess arises
from lens galaxy environments, since many of the $\theta\sim 3''$ lenses
appear to lie in groups or clusters. We extend this consideration
to all lensed quasars: Figure \ref{fig:sepdist_cas} shows the image
separation distribution for all lensed quasars and for the subset
whose lens galaxies lie in groups or clusters.  Lens galaxy 
environment is clearly correlated with image separation such that
lenses with larger separations tend to lie in groups and clusters.
The correlation is almost tautological at large separations (say
$\theta > 4''$), because it is hard to achieve a large deflection
angle without some environmental convergence. But the figure shows
a gradual increase of the fraction even at $\theta < 4''$.  This
trend provides further evidence that environmental convergence and
shear are important: it contradicts the prediction by \citet{keeton00}
that, if external convergence and shear are irrelevant, lenses in
dense environments should have a {\it smaller} mean image separation
than lenses in the field (because dense environments tend to have a
larger ratio of dwarf to giant galaxies than the field).

While these data certainly suggest a connection between environment
and lensing, the observational evidence is incomplete because systematic
surveys of lens environments have only just begun \citep[see][]{momcheva05}.
It is therefore valuable to undertake a theoretical analysis of how
environments affect lensing probabilities, and what range of environments
is expected for lenses with various image separations.  In this paper,
we compute the lensing effects of shear and convergence associated
with matter near the lens galaxy \citep[see also][for the effect of
groups on lens statistics]{moller02}.  We use a realistic model
for the distribution of galaxy environments, which is based on $N$-body
simulations and the halo occupation distribution and calibrated by
observations of galaxy--galaxy lensing and number counts of massive
elliptical galaxies \citep{dalal05}. For the lens
objects, we assume a singular isothermal sphere (SIS) because it is
both simple and a reasonable model for the density profile of lens
galaxies \citep[e.g.,][]{cohn01,treu02,koopmans03,rusin05}.  We neglect the
ellipticity of lens galaxies for simplicity, but note that the effects
of ellipticity are generally smaller than those of external shear
\citep{huterer05}.

This paper is organized as follows. Section \ref{sec:pdf} describes
the joint probability distribution function of convergence and shear
that we use throughout the paper.  Section \ref{sec:stat} reviews
the calculation of lensing probabilities and the image separation
distribution.  In \S \ref{sec:res} we examine the image separation
distribution from various viewpoints in order to understand how lens
environments affect the distribution.  Section \ref{sec:sum}
summarizes our main conclusions.  Throughout the paper we assume a
concordance cosmology with mass density $\Omega_M=0.3$, cosmological
constant $\Omega_\Lambda=0.7$, and dimensionless Hubble parameter
$h=0.7$.

\section{Joint PDF of convergence and shear}
\label{sec:pdf}

\begin{figure}
\begin{center}
 \includegraphics[width=1.0\hsize]{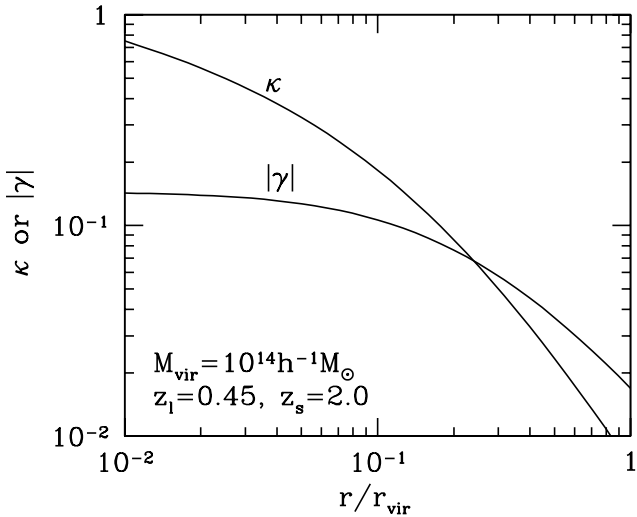}
\end{center}
\caption{Radial profile of convergence and shear in a typical
 massive dark matter halo. We assume an NFW halo with mass
 $M_{\rm vir} = 10^{14}\,h^{-1}\,M_\odot$ and concentration parameter
 $c=5.5$. The lens has redshift $z_l=0.45$, while the source has
 redshift $z_s=2$.
 \label{fig:kg_nfw}}
\end{figure}

We consider the probability distribution functions (PDFs) of
convergence and shear originating from lens galaxy environments,
using the model derived by \citet{dalal05}.  They placed galaxies
in an $N$-body simulation using a halo occupation distribution
calibrated to match number counts and 
tangential shear profiles measured for massive
elliptical galaxies in the SDSS \citep[see][]{sheldon04}, and
determined the convergence and shear at each galaxy position.
Figure 3 of their paper shows the resulting distributions.  The
convergence $\kappa_{\rm ext}$ and shear $|\gamma_{\rm ext}|$ have
similar means $\approx 0.03$, but very different tails: large
convergences ($\kappa_{\rm ext} \ga 0.1$) are more common than large
shears ($|\gamma_{\rm ext}| \ga 0.1$).  There are two reasons for
the difference.  First, for a typical group or cluster dark matter
halo, we expect that a large shear is always accompanied by an even
larger convergence.  Figure \ref{fig:kg_nfw} illustrates this idea
by showing the convergence and shear profiles for a dark matter halo
modeled with an NFW density profile \citep{navarro97}; in the central
region ($\kappa, |\gamma| \ga 0.1$), we see that $\kappa > |\gamma|$.
\footnote{More generally, for a power-law density profile
$\rho(r)\propto r^{-\eta}$, the convergence and shear are related as
$|\gamma|/\kappa=|\eta-1|/|3-\eta|$.  Therefore the argument holds
for different mass profiles, as long as they have inner profiles that
are shallower than the isothermal sphere ($\eta=2$). }
Second, convergence and shear sum differently when there are multiple
perturbers.  The probability distribution of convergence is highly
skewed, with a tail extending to large positive values but no
corresponding tail at large negative values.  In contrast, the
distributions of both shear components are symmetric about zero.
Thus, a large convergence provided by one perturber cannot be canceled
by contributions from other perturbers, while large shear values can
be canceled.  As we shall see, the longer tail of convergence is one
reason that convergence is more important than shear.

\begin{figure}
\begin{center}
 \includegraphics[width=1.0\hsize]{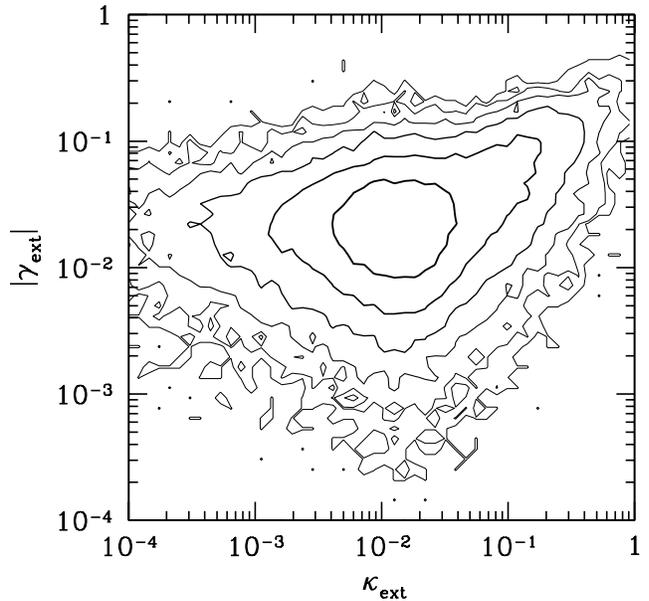}
\end{center}
\caption{Contours of the joint PDF $p(\kappa_{\rm ext},|\gamma_{\rm ext}|)$ 
 of external convergence and shear derived by \citet{dalal05}, assuming
 redshifts $z_l=0.45$ and $z_s=2$.  Contours are spaced by 0.5 dex, with
 thicker lines indicating higher probabilities.
\label{fig:kg_cont}}
\end{figure}

Figure \ref{fig:kg_nfw} also implies that convergence and shear are
not independent but rather correlated. Therefore in this paper we
consider the {\it joint} PDF of convergence and shear, which is shown
in Figure \ref{fig:kg_cont}. The correlation between $\kappa_{\rm ext}$
and $|\gamma_{\rm ext}|$ is quite evident, particularly for large
values.

Although the environment distribution was derived for a lens
redshift $z_l=0.45$, we assume little evolution in the elliptical
galaxy population over the range $0 \la z \la 1$
\citep[see][]{schade99,im02,ofek03,chae03b}.  Thus, we simply
extrapolate the surface mass density associated with environment
to all redshifts. Since convergence and shear depend on the surface
mass density in units of the critical density for lensing,
$\Sigma_{\rm crit} \propto D_{os}/(D_{ol} D_{ls})$, the redshift
dependence of the PDF for convergence and shear is straightforward.
The PDF was derived for galaxies with a velocity dispersion
$\sigma \sim 216$ km s$^{-1}$, which is typical for lenses with
image separation $\theta\sim 1''$, but we apply it to all galaxies.
In other words, we neglect any correlation between a galaxy's
velocity dispersion and environment.  However, we know that galaxies
with higher velocity dispersion will tend to reside in denser
environments; accordingly, our estimate of environmental effects on
wide-splitting lenses is strictly a lower limit.  
We exclude extreme cases with
$\kappa_{\rm ext}+|\gamma_{\rm ext}| \ge 1$, because those would be
categorized as lensing by groups/clusters rather than by a galaxy
perturbed by its environment.

\section{Lens Statistics with External Convergence and Shear}
\label{sec:stat}

We model each lens object as an SIS galaxy with external convergence
and shear.  The Einstein radius of an SIS with velocity dispersion
$\sigma$ is
\begin{equation}
 \theta_{\rm E}(\sigma)=4\pi\left(\frac{\sigma}{c}\right)^2
\frac{D_{ls}}{D_{os}},
\end{equation}
where $D_{os}$ and $D_{ls}$ are, respectively, angular diameter
distances from the observer and lens to the source.

The important properties of the lens are the image separation as
a function of source position, the lensing cross section, and the
magnification bias.  We define the image separation $\theta$ to be
the maximum distance between multiple images, which is a well-defined
and observable quantity.  It is useful to define the normalized
separation $\hat{\theta}$ such that
\begin{equation}
 \theta(\mathbf{u}) = \theta_{\rm E}\,\hat{\theta}(\mathbf{u}).
\end{equation}
Note that $\hat{\theta}(\mathbf{u})=2$ for all $\mathbf{u}$ in the
absence of convergence and shear.

We combine the cross section and magnification bias to compute the
``biased cross section'',
\begin{equation}
 BA = \int \frac{\Phi(L/\mu)}{\Phi(L)}\,d\mathbf{u}
 = \int \mu^{\beta-1}\,d\mathbf{u},
\end{equation}
where the integral is over the multiply-imaged region of the
source plane, and we assume a power law source luminosity function
$\phi_L(L)\propto L^{-\beta}$.  Unless otherwise specified, we adopt
the source luminosity function of the CLASS survey, $\beta=2.1$
\citep{myers03}. We take $\mu$ to be the total magnification of
all images, which is appropriate for surveys in which lenses are
identified from high-resolution follow-up observations of unresolved
targets (including both CLASS and SDSS).  We choose coordinates such
that the source position $\mathbf{u}$ is in units of $\theta_{\rm E}$,
which means that $BA$ is naturally (and conveniently) in units of
$\theta_{\rm E}^2$.

We compute image separations and biased cross sections using the
public software {\it gravlens} by \citet{keeton01b}.  For each set
of ($\kappa_{\rm ext}$, $|\gamma_{\rm ext}|$) we place $10^5$ random
sources in the smallest circle enclosing the lensing caustics.  We
solve the lens equation to find the images, and then compute the
image separation and total magnification.  Finally, we sum over the
multiply-imaged sources to compute the biased cross section.

The next step is to integrate over lens galaxy populations to obtain
the total lensing probability. This involves integrating over
appropriate distributions of galaxy masses (or velocity dispersions),
redshifts, and environments:
\begin{eqnarray}
 P&=&\int dz_l\frac{c\,dt}{dz_l}(1+z_l)^3\int d\kappa_{\rm ext} d
 |\gamma_{\rm ext}|\,p(\kappa_{\rm ext},|\gamma_{\rm ext}|)\nonumber\\
 &&\times\int d\sigma \frac{dn}{d\sigma}
  \left(D_{ol}\theta_{\rm E}\right)^2 BA .
\label{p_tot}
\end{eqnarray}
Here $D_{ol}$ is the angular diameter distance from the observer to
the lens, and $p(\kappa_{\rm ext}, |\gamma_{\rm ext}|)$ is the joint
PDF of external convergence and shear from \S \ref{sec:pdf}.  We
specify the distribution of galaxy velocity dispersions using the
velocity function $dn/d\sigma$ of early-type galaxies determined
from $\sim$30,000 galaxies at $0.01<z<0.3$ in the SDSS
\citep{sheth03,mitchell05}.

The image separation distribution of lensed quasars can then be
obtained by differentiating equation (\ref{p_tot}):
\begin{eqnarray}
 \frac{dP}{d\theta}&=&\int dz_l\frac{c\,dt}{dz_l}(1+z_l)^3 \int 
d\kappa_{\rm ext} d|\gamma_{\rm ext}| \,p(\kappa_{\rm ext},|\gamma_{\rm
ext}|)\nonumber \\ && \times \int d\hat{\theta}\frac{1}{\hat{\theta}}
 \frac{d\sigma}{d\theta_{\rm
 E}}\frac{dn}{d\sigma}\left(D_{ol}\theta_{\rm E}\right)^2
\frac{d(BA)}{d\hat{\theta}}\\ 
&=&\int dz_l\frac{c\,dt}{dz_l}(1+z_l)^3 \int d\kappa_{\rm ext}
 d|\gamma_{\rm ext}| \,p(\kappa_{\rm ext},|\gamma_{\rm ext}|)\nonumber \\ 
 && \times \int d\hat{\theta} \frac{\sigma}{2\theta} \frac{dn}{d\sigma} 
 \frac{\left(D_{ol}\theta\right)^2}{\hat{\theta}^2}\frac{d(BA)}{d\hat{\theta}}. 
\label{dpdt}
\end{eqnarray}
We fix the source redshift to $z_s=2.0$ which is a typical redshift
for lensed quasars.

\section{Results}
\label{sec:res}

\subsection{Dependence on convergence and shear}
\label{sec:conshe}

\begin{figure}
\begin{center}
 \includegraphics[width=1.0\hsize]{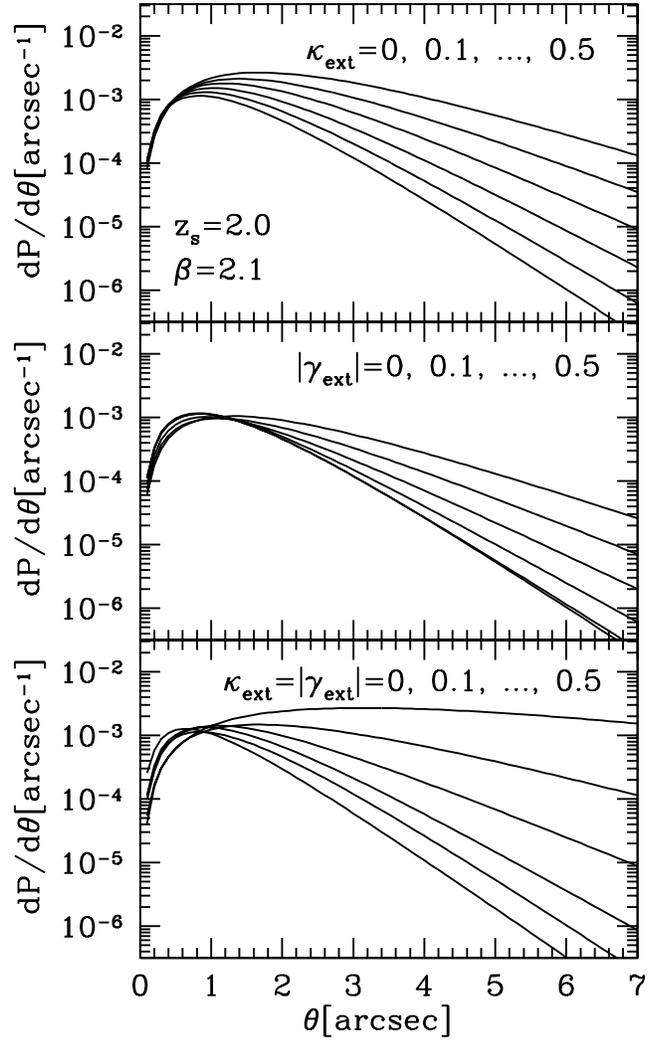}
\end{center}
\caption{Effects of external convergence and shear on the predicted
 image separation distribution. Here we assume certain fixed values of
 convergence and/or shear instead of taking account of full joint
 PDF. Top: We vary the convergence $\kappa_{\rm ext}$ from 0 to 0.5
 (bottom to top), while the shear is fixed at $|\gamma_{\rm ext}|=0$.
 Middle: We vary the shear $|\gamma_{\rm ext}|$ from 0 to 0.5 (bottom
 to top at $\theta \ga 1''$), while the convergence is fixed at
 $kappa_{\rm ext}=0$.  Bottom: We set $\kappa_{\rm ext}=|\gamma_{\rm ext}|$
 and change the value from 0 to 0.5 (bottom to top at $\theta \ga 1''$).
\label{fig:sepdist}}
\end{figure}

Before presenting results that account for full distribution of
convergence and shear, it is useful to study how fixed values of
$\kappa_{\rm ext}$ and/or $|\gamma_{\rm ext}|$ affect the image
separation distribution.  These results are shown in
Figure \ref{fig:sepdist}.

Convergence and shear both modify the shape of image separation
distribution, particularly at relatively large image separations.
For instance, at $\theta=5''$ a convergence $\kappa_{\rm ext}=0.2$
increases the lensing probability by one order of magnitude, and
a shear of $|\gamma_{\rm ext}|=0.2$ enhances it by a factor 2--3.
For the same values, convergence clearly has much more effect on
lensing probabilities than shear.  Combined with the fact that 
large values of convergence are more likely to occur than large
values of shear (see \S \ref{sec:pdf}), we expect that convergence
has the greater impact on image separation distributions.

The results can be understood as follows. First, external convergence 
magnifies the image plane with respect to the source plane, which
increases the image separation by a factor $(1-\kappa_{\rm ext})^{-1}$.
Second, while external convergence does not affect the caustics, it
does change the biased cross section by giving additional magnification
to the images. Specifically, the biased cross section increases by a
factor $(1-\kappa_{\rm ext})^{-2(\beta-1)}$ \citep[see][]{keeton04}.
As a result of these two effects, convergence shifts the image separation
distribution up and to the right in Figure \ref{fig:sepdist}.  The
increase is significant where the image separation distribution is a
deceasing function ($\theta \ga 1''$), but negligible where it is an
increasing function ($\theta \ll 1''$). By contrast, shear mainly
broadens the image separation distribution. It modestly enhances the
lensing probability at $\theta \ga 1''$, and suppresses it at
$\theta \la 1''$.

\subsection{Full result}

\begin{figure}
\begin{center}
 \includegraphics[width=1.0\hsize]{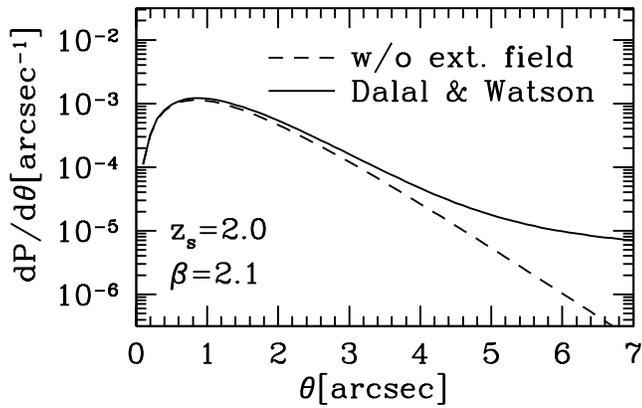}
\end{center}
\caption{Image separation distributions with ({\it solid}) and without
 ({\it dashed}) environmental convergence and shear, calculated from
 equation (\ref{dpdt}). We now use the joint distribution of external
 convergence and shear from Figure \ref{fig:kg_cont}.
\label{fig:sepdist_full}}
\end{figure}

\begin{figure}
\begin{center}
 \includegraphics[width=1.0\hsize]{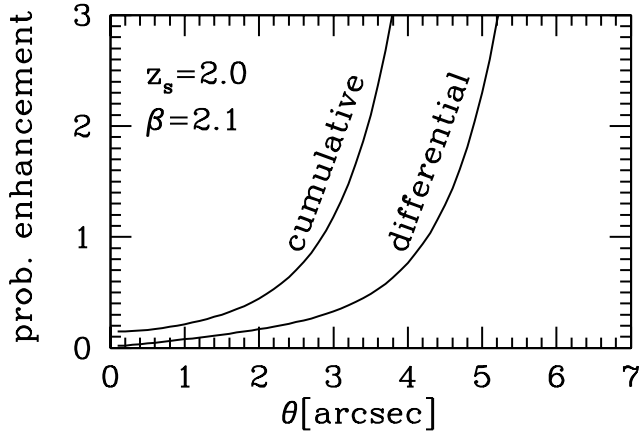}
\end{center}
\caption{Enhancement of the lensing probability as a function of
image separation.  The enhancement is defined to be $P_{\rm ext}/P_0 - 1$,
where $P_{\rm ext}$ is the probability with the external field while
$P_0$ is the probability without.  We show the enhancement for both
the differential lensing probability $dP/d\theta$ and the cumulative
probability $P(>\theta)$.  \label{fig:sepdist_rat}}
\end{figure}

We are now ready to consider the image separation distribution with
the full joint PDF of convergence and shear, which is shown in
Figure \ref{fig:sepdist_full}.  The enhancement of the lensing
probability, which is plotted in Figure \ref{fig:sepdist_rat}, is
small at $\theta \sim 1''$, but significant at larger image
separations: the increase is $\sim$30\% at $\theta=3''$ and
$\sim$200\% at $\theta=5''$. Thus, lens galaxy environments are
very important when we discuss the shape of the image separation
distribution. Indeed, they can account for the excess of lensing
probabilities reported by \citet{oguri05}.

In Figure \ref{fig:sepdist_rat}, we also plot the enhancement of the
cumulative lens probability $P(>\theta)$. We find that the enhancement
of the total lensing probability integrated over the all image
separation is $\sim 15\%$. This enhancement is equivalent to that
obtained by shifting the cosmological constant by
$\Delta\Omega_\Lambda\sim 0.05$, therefore it cannot be ignored in
accurate determination of cosmological parameters with lens statistics.

\begin{figure}
\begin{center}
 \includegraphics[width=1.0\hsize]{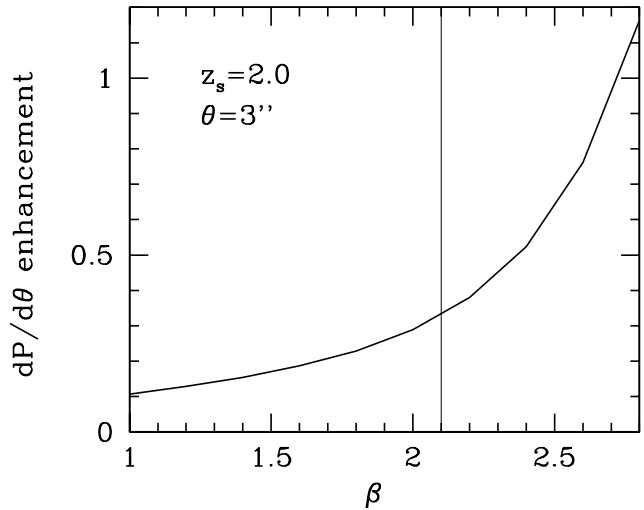}
\end{center}
\caption{Enhancement of the lensing probability, computed at an image
 separation $\theta=3''$, as a function of the slope $\beta$ of the
 source luminosity function.  The vertical line indicates the fiducial
 value we adopt in this paper ($\beta=2.1$, as appropriate for the
 CLASS lens survey). 
\label{fig:beta}} 
\end{figure}

As discussed above, some of the effects of convergence and shear are
mediated by magnification bias: convergence directly magnifies images,
while shear generates and lengthens the tangential caustic and thus
increases the cross section for high magnifications.  The enhancement
of the lensing probability is therefore expected to depend on the
source luminosity function. Indeed, Figure \ref{fig:beta} shows that
the enhancement at $\theta=3''$ increases strongly with the luminosity
function slope $\beta$ (recall $\Phi \propto L^{-\beta}$).  This
result implies that bright optical lensed quasars with large image
separations are more likely to lie in dense environments.

\subsection{What enhances the lensing probability?}

\begin{figure}
\begin{center}
 \includegraphics[width=1.0\hsize]{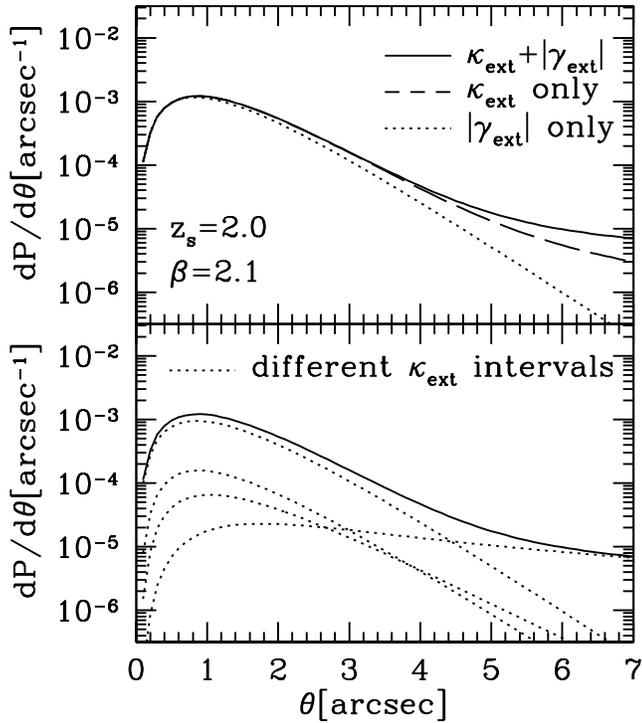}
\end{center}
\caption{Top: Image separation distributions that include only
 convergence ({\it dashed}), only shear ({\it dotted}), or both
 ({\it solid}). Bottom: Contributions to the image separation
 distribution from different $\kappa_{\rm ext}$ intervals.  From
 top to bottom (at $\theta \sim 1''$), the dotted curves correspond
 to $\kappa_{\rm ext}<0.05$, $0.05<\kappa_{\rm ext}<0.15$,
 $0.15<\kappa_{\rm ext}<0.35$, and $0.35<\kappa_{\rm ext}$. The full
 image separation distribution is shown with the solid line for
 reference. 
\label{fig:sepdist_dec}} 
\end{figure}

Dissecting the results further helps us understand the changes to
the image separation in more detail.  As discussed in
\S \ref{sec:conshe}, we believe that convergence is more important
than shear because of its stronger effect on the lensing probability
(see Fig.~\ref{fig:sepdist}) and the longer tail to high
$\kappa_{\rm ext}$ in the joint PDF (see Fig.~\ref{fig:kg_cont}). To
check this hypothesis, we project the joint PDF to a distribution of
convergence or shear alone (the same distributions shown in Fig.~3
\citealt{dalal05}) and recompute the image separation distribution.
The results, shown in the upper panel of Figure \ref{fig:sepdist_dec},
confirm that the enhancement of the lensing probability is in fact
driven by convergence.

In that case, it is useful to understand what values of
$\kappa_{\rm ext}$ contribute most to the lensing probability. We
see this by decomposing the image separation distribution into
contributions from different $\kappa_{\rm ext}$ intervals, as shown
in the bottom panel of Figure \ref{fig:sepdist_dec}. The
interpretation depends on the image separation. At small image
separations ($\theta<3''$), most lenses are expected to have small
convergences ($\kappa_{\rm ext} \la 0.05$), and larger and larger
convergences are more and more rare.  At intermediate separations
($3'' \la \theta \la 4''$), the largest convergences
($\kappa_{\rm ext} \ga 0.35$) become more important than
intermediate values.  Finally at large separations ($\theta \ga 5''$),
the largest convergences become dominant.  In other words, while
dense environments with large convergences may be rare, nearly all
of the largest separation lenses will be found there.

\subsection{Environments of lens galaxies}

\begin{figure}
\begin{center}
 \includegraphics[width=1.0\hsize]{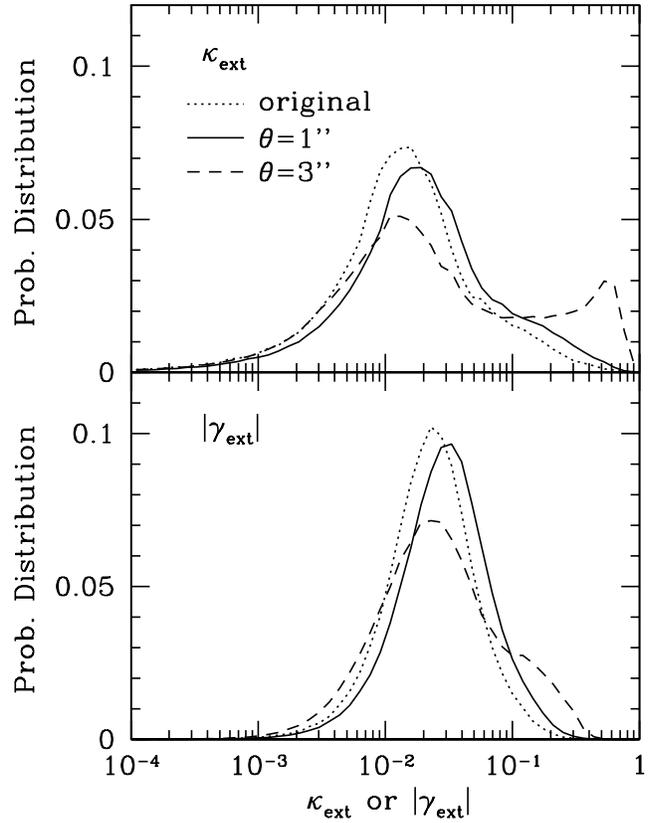}
\end{center}
\caption{Posterior distributions of convergence ({\it top}) and shear
 ({\it bottom}), taking into account lensing probabilities.  Distributions
 for lenses with image separations $\theta=1''$ ({\it solid}) and $3''$
 ({\it dashed}) ares shown. The original (unweighted) distributions at
 $z_l=0.45$ are shown by dotted lines for reference.
\label{fig:kg_hist_lens}}
\end{figure}

Figure \ref{fig:sepdist_dec} suggests that the typical environments
of lenses are very sensitive to image separation.  Put another way,
the distribution of convergence and shear for actual lens galaxies
will differ from the distribution shown in Figure \ref{fig:kg_cont}
(which applies to normal elliptical galaxies), by a factor of the
lensing probability.  To quantify this effect, we derive the
posterior PDF of convergence and shear for lenses at fixed image
separations of $\theta=1''$ or $3''$.  Figure \ref{fig:kg_hist_lens}
shows that the distributions differ from each other, and from the
original (unweighted) distributions.  The distributions for lenses
with $\theta=1''$ are similar to the original distributions, just
shifted to slightly larger values.  However, the distributions for
lenses with $\theta=3''$ show particularly large increases in the
tail of high convergence or shear values.

\begin{figure}
\begin{center}
 \includegraphics[width=1.0\hsize]{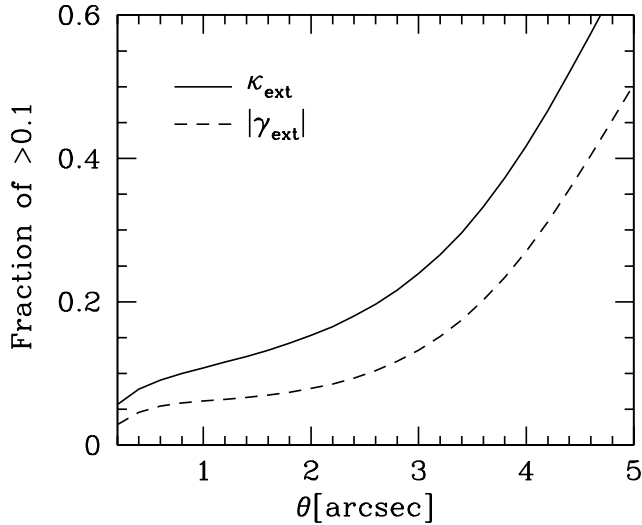}
\end{center}
\caption{Fraction of lenses with ``strong'' environments (defined by
 $\kappa_{\rm ext}$ or $|\gamma_{\rm ext}| > 0.1$), as a function of
 the image separation $\theta$.
\label{fig:frac}}
\end{figure}

To see this more clearly, in Figure \ref{fig:frac} we show the
fraction of lenses with a ``strong'' environment (defined to be
$\kappa_{\rm ext}$ or $|\gamma_{\rm ext}| > 0.1$), as a function
of image separation $\theta$. The fraction increases monotonically
as the image separation grows. For instance, at $\theta=1''$ the
fraction of lenses with a strong convergence is 11\%, while the
fraction with a strong shear is 6\%.  By $\theta=3''$ the fractions
have increased to 24\% and 13\%, respectively.  (For comparison,
the fractions of non-lens elliptical galaxies with strong convergence
or shear are just 6\% and 3\%; see \citealt{dalal05}.) In other
words, the distribution of lens galaxy environments is a strong
function of image separation, and this effect should be taken into
account when comparing observed environments with theoretical
predictions.  The increasing fraction of ``strong'' environments
with image separation is qualitatively consistent with the data
shown in Figure \ref{fig:sepdist_cas}.  
 
\section{Summary and Discussion}
\label{sec:sum}

We have studied how the convergence and shear from lens galaxy
environments affect lens statistics, in particular the distribution
of lens image separations.  We find that the external field enhances
the lensing probability, especially at large image separations, and
the effect increases with the slope of the source luminosity function.
We argue that the enhancement is driven mainly by the external
convergence, which has been neglected in previous studies of this
sort.  Our results mesh with those of \citet{keeton04} to indicate
that it is essential to include convergence from the environment
in order to obtain correct results from both lens modeling and
lens statistics. For example, since the external field changes the
shape of the image separation distribution rather than its overall
amplitude, it will bias attempts to constrain the velocity dispersion
function of early-type galaxies using the lens image separation
distribution \citep[e.g.,][]{chae05}.  

The environmental boost in the lensing probability depends on the
image separation, which means that the posterior distribution of
lens environments does as well.  The fraction of lenses with a
``strong'' environment (a convergence or shear larger than 0.1, say)
increases monotonically with image separation.  Large separation
lenses are more likely to be found in dense environments.  This bias
will be important when comparing the observed distribution of lens
environments to theoretical predictions \citep[e.g.,][]{momcheva05}.

One puzzle remains: even when we take the separation/environment
correlation into account, we find that the predicted fraction of
lens galaxies with a large shear $|\gamma_{\rm ext}| > 0.1$ is
$\sim$10\%.  However, models of most observed 4-image lenses
require shears of $\sim$0.1 or larger \citep{keeton97}.  While
4-image lenses are certainly biased toward dense environments
(see \citealt{holder03} for a theoretical perspective, and
\citealt{momcheva05} for intriguing observational results), it does
not appear that the bias is strong enough to reconcile our
predicted shear distribution with the ``observed'' values.  At
present, it is not clear whether the problem lies with the lens
models or the predicted environment distributions.  If the latter,
then our conclusions regarding the importance of lens galaxy
environments should clearly be revisited.  

It is also possible that relaxing some simplifying assumptions in
our analysis will affect the quantitative results.  The most
important simplification is the assumption that all galaxies have
the same distribution of convergence and shear.  As shown in
\citet{sheldon04}, however, the tangential shear signal on scales
$\lesssim 1\,h^{-1}$ Mpc increases with increasing velocity
dispersion, suggesting that more massive ellipticals reside in denser
environments than less massive ellipticals.  We expect this
effect to enhance the importance of environment for
wide-splitting lenses relative to the estimates presented here.  

Another simplification was our neglect of lens galaxy ellipticity.
The effects of ellipticity are somewhat similar to those of shear
\citep[see][]{huterer05}, which has a modest effect on the lensing
probability.  Therefore we expect that our results would not change
significantly with the addition of ellipticity.  Nevertheless,
since ellipticity is important in predictions of the relative
numbers of 4-image and 2-image lenses, it will need to be considered
more carefully when comparing observed and predicted environment
distributions for 4-image lenses alone.

It seems to be too early to say whether there is conflict between
the observed and predicted distribution of lens galaxy environments.
But it is an intriguing question that deserves to be studied both
observationally and theoretically.

\section*{Acknowledgments}
We thank the referee, David Rusin, for helpful comments and
suggestions. M.\ O.\ is supported by JSPS through JSPS Research
Fellowship for Young Scientists.   N.\ D.\ acknowledges the support of
NASA through Hubble Fellowship grant \#HST-HF-01148.01-A awarded by
the Space Telescope Science Institute, which is operated by the
Association of Universities for Research in Astronomy, Inc., for NASA,
under contract NAS 5-26555. 


\label{lastpage}

\end{document}